\documentclass[sigconf]{acmart}
\usepackage{graphicx}
\usepackage{caption}
\usepackage[ruled,linesnumbered]{algorithm2e}
\usepackage{balance}  
\usepackage{amsmath}
\usepackage{tabularx}
\usepackage{multirow}
\usepackage[utf8]{inputenc}
\usepackage{url}
\PassOptionsToPackage{hyphens}{url}\usepackage{hyperref}
\usepackage{tabto}
\newcolumntype{d}[1]{D{.}{.}{#1}}

\begin{document} 

\title{Bio-SODA: Enabling Natural Language Question Answering over Knowledge Graphs without Training Data}

\author{Ana Claudia Sima}
\affiliation{\institution{SIB Swiss Institute of Bioinformatics}
\city{Lausanne}
\country{Switzerland}}
\email{Ana-Claudia.Sima@sib.swiss}

\author{Tarcisio Mendes de Farias}
\affiliation{\institution{SIB Swiss Institute of Bioinformatics}\city{Lausanne}\country{Switzerland}}
\affiliation{\institution{University of Lausanne}\city{Lausanne}
\country{Switzerland}}
\email{Tarcisio.Mendes@sib.swiss}

\author{Maria Anisimova}
\affiliation{
\institution{Zurich University of Applied Sciences}\city{W\"{a}denswil}
\country{Switzerland}}
\affiliation{\institution{SIB Swiss Institute of Bioinformatics}
\city{Lausanne}
\country{Switzerland}}
\email{Maria.Anisimova@zhaw.ch}

\author{Christophe Dessimoz}
\affiliation{\institution{University of Lausanne}\city{Lausanne}
\country{Switzerland}}
\affiliation{\institution{University College London}\city{London}
\country{UK}}
\affiliation{\institution{SIB Swiss Institute of Bioinformatics}\city{Lausanne}
\country{Switzerland}}
\email{Christophe.Dessimoz@unil.ch}

\author{Marc Robinson-Rechavi}
\affiliation{\institution{University of Lausanne}\city{Lausanne}
\country{Switzerland}}
\affiliation{\institution{SIB Swiss Institute of Bioinformatics}\city{Lausanne}
\country{Switzerland}}
\email{Marc.Robinson-Rechavi@unil.ch}

\author{Erich Zbinden}
\affiliation{\institution{Zurich University of Applied Sciences}\city{W\"{a}denswil}
\country{Switzerland}}
\email{Erich.Zbinden@zhaw.ch}

\author{Kurt Stockinger}
\affiliation{\institution{Zurich University of Applied Sciences}\city{Winterthur}
\country{Switzerland}}
\email{Kurt.Stockinger@zhaw.ch}

\setcopyright{acmcopyright}
\copyrightyear{2021}
\acmYear{2021}

\acmConference[SSDBM '21]{SSDBM '21: 33rd International Conference on Scientific and Statistical Database Management}{July 06-07, 2021}{}
\acmISBN{}
\acmBooktitle{}
\begin{abstract}
The problem of natural language processing over structured data has become a growing research field, both within the relational database and the Semantic Web community, with significant efforts involved in question answering over knowledge graphs (KGQA). However, many of these approaches are either specifically targeted at {\it open-domain} question answering using DBpedia, or require \textit{large training datasets} to translate a natural language question to SPARQL in order to query the knowledge graph. Hence, these approaches often cannot be applied directly to complex {\it scientific datasets} where no prior training data is available.

In this paper, we focus on the challenges of natural language processing over knowledge graphs of scientific datasets. In particular, we introduce Bio-SODA, a natural language processing engine that does not require training data in the form of question-answer pairs for generating SPARQL queries. Bio-SODA uses a generic graph-based approach for translating user questions to a ranked list of SPARQL candidate queries. Furthermore, Bio-SODA uses a novel ranking algorithm that includes node centrality as a measure of relevance for selecting the best SPARQL candidate query. Our experiments with real-world datasets across several scientific domains, including the official \textit{bioinformatics} Question Answering over Linked Data (QALD) challenge, as well as the CORDIS dataset of European projects, show that Bio-SODA outperforms publicly available KGQA systems  by an F1-score of least 20\% and by an even higher factor on more complex bioinformatics datasets. 

\end{abstract}

\ccsdesc[500]{Information systems~Graph-based database models}
\ccsdesc[500]{Information systems~Retrieval models and ranking}

\keywords{Question Answering, Knowledge Graphs, Ranking}

\maketitle
\section{Introduction}
\label{sec:Introduction}

The problem of natural language processing over structured data has gained significant traction, both in the Semantic Web community -- with a focus on answering natural language questions over RDF graph databases \cite{diefenbach2018towards, zheng2018question, vakulenko2019message} -- and in the relational database community, where the goal is to answer questions by finding their semantically equivalent translations to SQL \cite{li2014constructing, li2016understanding, saha2016athena,brunner2021valuenet}. Significant research efforts have been invested in particular in \textit{open-domain} question answering over knowledge graphs. These efforts often use the DBpedia and/or Wikidata knowledge bases, that are composed of structured content from various Wikimedia projects such as Wikipedia. A growing ecosystem of tools is therefore becoming available for solving subtasks of the KGQA problem, such as entity linking \cite{sakor2019falcon, ferragina2010tagme, mendes2011dbpedia, olieman2014entity} or query generation \cite{zafar2018formal}. However, most of these tools are specifically targeted at question answering over DBpedia \cite{singh2018no}, which casts doubts on their applicability to other contexts, such as for scientific datasets. 

On the one hand, encouraged by the recent success of machine learning methods, several new benchmarks for training and evaluating KGQA systems have been published \cite{trivedi2017lc, dubey2019lc}. On the other hand, most of the existing datasets are synthetic (\textit{i.e.}, not based on real query logs) and generally limited to DBpedia or Wikidata, which may not be representative of knowledge graphs for scientific datasets. 

For example, one of the major question answering datasets over DBpedia, LC-Quad \cite{trivedi2017lc}, as well as its updated version, LC-Quad 2.0 \cite{dubey2019lc}, include only {\it simple multi-fact questions} that connect at most two facts. In other words, these queries cover at most two or three triple patterns, with a query graph spanning a maximum of two hops, whereas real-world questions tend to be much more complex. In particular, a study of SPARQL query logs \cite{bonifati2019analytical} across multiple knowledge graphs, including DBpedia, has shown that a significant fraction of real-world queries have 10 triple patterns or more. It therefore remains unclear whether existing training sets can serve as representative for real-world natural language processing engines over knowledge graphs in general. 

All in all, an important unknown still remains as to how many of the lessons learned in question answering over DBPedia can be easily applied to querying scientific datasets. In these domains, an equivalent ecosystem of tools is not readily available. As a consequence, data access and retrieval remain challenging for domain experts who are not familiar with structured query languages, nor with the data models of each scientific dataset that they use. 

\begin{figure}[h]
    \centering
    \includegraphics[width=0.8\columnwidth]{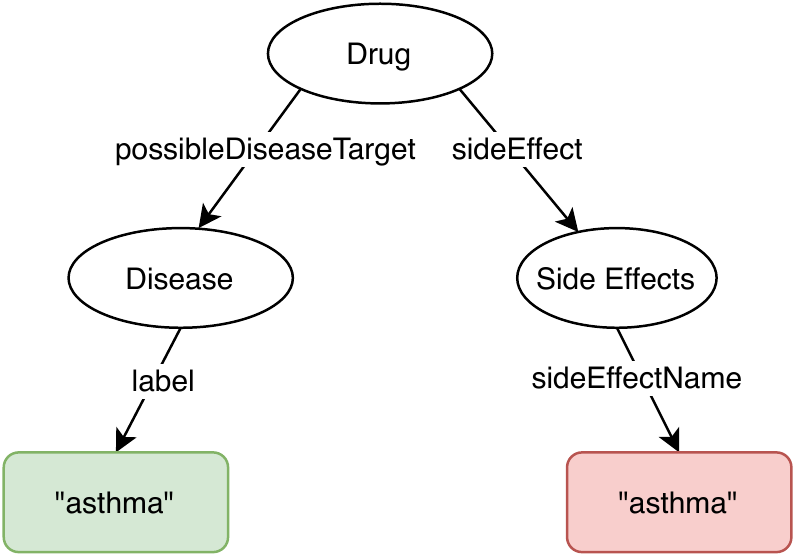}
    \caption{Illustrative data model, simplified from the QALD4 benchmark datasets \cite{unger2014question}. Consider the following question: \textit{``Which drugs are used for asthma?"}. In the QALD4 dataset, "asthma" appears as both a disease instance (shown in green), as well as a side effect (shown in red). The second interpretation describes drugs that can \textit{trigger} asthma symptoms. Therefore, it is the opposite of the user's intended question. However, the predicate \textit{used for} in the question cannot be easily linked to either of the properties indicated through arrows in the image. Due to ambiguity, the question is difficult to translate correctly in the absence of external knowledge, without relying on training data (inferring that \textit{used for} implies drug targeting disease).} 
    \label{fig:Toy Data Model}
\end{figure}

To illustrate the general problem of natural language processing over knowledge graphs, consider the simple data model in Figure \ref{fig:Toy Data Model}. Here we see that a drug could be a {\it possible disease target} for asthma (left branch), as well as potentially having {\it side effects} such as triggering asthma symptoms (right branch). Now consider the following natural language question: \textit{"Which drugs are used for asthma?"}. Note that our knowledge graph has no concept or property called \textit{used for}. Hence, this question cannot be easily translated without relying on external knowledge (\textit{e.g.} training data), given that \textit{used for} cannot be directly mapped to either of the two properties (\textit{possibleDiseaseTarget} or \textit{sideEffect}) shown in the figure. However, node centrality metrics, such as the PageRank score of nodes in the knowledge graph, can help capture "common sense" knowledge, \textit{e.g.}, that \textit{asthma} is more commonly a \textit{Disease}, rather than a \textit{Side Effect}. 

As a step towards bridging the current gap in natural language processing for knowledge graphs of scientific datasets, we introduce Bio-SODA, a system designed to answer \textit{natural language questions} across  knowledge graphs where \textit{no prior training data} is available. Bio-SODA relies on a generic graph-based approach in order to translate natural language questions into SPARQL queries. Furthermore, Bio-SODA is designed to \textit{compensate for incompleteness in the data}---either due to missing schema information or, to some extent, due to missing labels. Although these situations should not occur when following ontology engineering best practices for representing data in RDF, our experience in working with real-world datasets shows that these problems are frequent in practice. 

We make our prototype implementation available open-source\footnote{Code at \url{https://github.com/anazhaw/Bio-SODA}}. We also make available a live demo of Bio-SODA online\footnote{See demo at \url{http://biosoda.expasy.ch/welcome/}}, where each of the datasets considered in this paper can be queried. The prototype enables both keyword search, as well as full question answering in English. We chose bioinformatics as our \textit{primary} target domain, motivated by the rapid growth of publicly available RDF data in this scientific domain. Specifically, around 8\% of the Linked Open Data Cloud originates from the Life Sciences \cite{hasnain2017biofed}. For the purpose of evaluating our system, we use several real-world datasets stemming from different domains. For example, we use the last bioinformatics question answering challenge released as part of the official Question Answering on Large Databases (QALD) series, namely the QALD4 biomedical task \cite{unger2014question}. Importantly, to-date there is no sufficiently large training dataset of questions and corresponding SPARQL queries to enable the use of machine learning approaches for end-to-end Question Answering in the biomedical field.  Finally, to demonstrate the generalizability of Bio-SODA to other domains, we also apply our system to an entirely different context, outside bioinformatics, namely on the CORDIS dataset describing European Union (EU) funded research projects\footnote{https://cordis.europa.eu/projects}. This dataset is also used in the EU-project INODE (Intelligent Open Data Exploration) \cite{ameryahia2021inode}.


This paper makes the following {\bf contributions}:
\begin{itemize}
    \item  We introduce Bio-SODA---a novel natural language processing engine over knowledge graphs that {\it does not require prior training data} (question-answer pairs) for translating natural language questions into SPARQL.
    \item We define a novel ranking algorithm for selecting the best automatically generated SPARQL statements in response to a given natural language question. The ranking algorithm combines {\it syntactic and semantic similarity}, as well as {\it node centrality} in the knowledge graph. Many existing question answering systems either rely on simple metrics for ranking, such as the length of the answer query graph  \cite{saha2016athena}, or require extensive training data in order to learn a ranking function \cite{maheshwari2019learning}. To the best of our knowledge, our approach is the first to take into account all three factors (syntactic and semantic similarity, as well as node centrality) for ranking queries.
    \item Our experiments on various real-world datasets show that Bio-SODA outperforms state-of-the-art KGQA systems by 20\% on the F1-score using the official QALD4 biomedical benchmark and by an even higher factor on the more complex bioinformatics dataset. 
\end{itemize}

The paper is structured as follows: Section \ref{Related Work} places our contribution in the context of the related work. In Section \ref{Problem Statement} we introduce  some of the challenges of natural language processing over RDF-based knowledge graphs. In Section \ref{Section: Question Answering Pipeline} we explain the high level architecture of Bio-SODA through a concrete example from the biomedical domain. We present the detailed system architecture of Bio-SODA in Section \ref{sec:System Architecture}. Next, we describe the datasets used for evaluation, their specific challenges and the results obtained, in Section \ref{Evaluation}. In Section \ref{sec:lessons_learned} we discuss lessons learned from building a natural language processing system for real-world domain datasets.  We outline  directions for future work in Section \ref{Conclusions}.

\section{Related Work}
\label{Related Work}
The problem of natural language processing and question answering over structured data has been well-studied in recent years, with a growing number of published systems, particularly in open-domain question answering. Recent surveys on natural language interfaces to databases include \cite{affolter2019comparative, chakraborty2019introduction}. However, in this paper we focus on natural language interfaces to \textit{RDF graph databases or RDF-based knowledge graphs}. Natural language interfaces to relational databases are outside the scope of this paper.

In parallel, the biomedical field has seen a growth of dedicated systems for question answering. Examples include GFMed \cite{marginean2017question} and Pomelo \cite{hamon2014description} -- the two highest ranked systems in the QALD4 biomedical challenge -- as well as more recent systems \cite{hamon2017querying}. However, these are generally considered expert systems, with lower generalizability to other domains, given that they extensively rely on manually handcrafted rules and domain expertise.

Our work aims to bridge the gap between the two parallel efforts by solving the \textit{common case} in a domain-independent manner. For this, Bio-SODA relies on a generic graph-based approach in order to generate a ranked list of candidate SPARQL queries from a given question. We enable the addition of custom rules only for \textit{special cases} when needed. 

Many recent KGQA systems \cite{vakulenko2019message, diefenbach2018towards} have been evaluated using the LC-Quad benchmark of 5000 questions over DBpedia \cite{trivedi2017lc}. Although this benchmark is an important step forward, particularly for enabling  machine learning approaches, it does not include complex multi-hop questions, which makes it unclear how the results would generalize to this case. For example, at the time of writing, the current publicly available implementation of the SPARQL query generation system SQG \cite{zafar2018formal}, would not work for complex question answering on a new knowledge graph without significant changes to the code base, as it targets question answering over DBpedia and more specifically in the format required by the LC-Quad benchmark. 

More recent KGQA systems, such as  \cite{diefenbachqanswer, vakulenko2019message}, support multiple knowledge graphs, but are limited to queries with a complexity of at most three triple patterns. Similarly, existing end-to-end QA systems, based on machine learning approaches, such as \cite{lukovnikov2017neural}, can only handle simple questions. These approaches have the added drawback that they only generate a single answer, as opposed to multiple candidates. Furthermore, end-to-end approaches suffer from the lack of explainability, which makes it challenging for users to validate the correctness of the result. Explainability in this context has therefore become an active area of research, with solutions proposed including translating back structured queries into natural language sentences \cite{deutch2020explaining, ngonga2013sorry, kokkalis2012logos} or summarizing the entities in the results \cite{diefenbach2018pagerank}.


Disambiguation is one of the major tasks of question answering systems. One possible solution for this is to limit the interface to a controlled natural language and involve the user in constructing questions from the available building blocks. Sparklis \cite{ferre2017sparklis} is a query building system that enables answering controlled natural language questions over knowledge graphs out-of-the-box. However, this process is manual and therefore time-consuming, which makes it less convenient than a true natural language interface.

One of the systems closest to ours is the KG-agnostic WDAqua-core1 \cite{diefenbach2018towards}. The system supports multiple knowledge bases in several languages. However,  the system is only available as a demo. Although the authors mention that node relevance can in principle be taken into account for ranking, it is not clear whether the approach was used in the evaluation or whether the ranking function was learned based on training data. 

\section{Challenges of Natural Language Processing over Knowledge Graphs}
\label{Problem Statement}

In this section we summarise some of the challenges of natural language processing over knowledge graphs, focusing on scientific knowledge graphs, which shape the architecture of the Bio-SODA system (described in Sections \ref{Section: Question Answering Pipeline} and \ref{sec:System Architecture}).

\begin{itemize}
    \item \textbf{Lack of training data.}
    
    For many scientific knowledge graphs there is no sufficiently long and diverse log of queries in order to derive a representative training set for a machine learning-based solution. So far, existing training corpora have proven costly to construct \cite{trivedi2017lc}, with the added drawback that any semi-automatically generated dataset risks compiling a set of question-answer pairs that are non-representative for the information needs of real users of the KGQA system, \textit{e.g.} domain experts.
    
    
    
    
    
    \item \textbf{Rule-based approaches perform well, but are costly to build and maintain.}
    
    So far, state-of-the-art solutions for question answering over generic RDF-based knowledge graphs have been mostly rule-based systems, relying on manually handcrafted rules. For example, GFMed \cite{marginean2017question} and Pomelo \cite{hamon2014description}, the top 2 ranked systems in the QALD4 biomedical challenge, have achieved very good results in the challenge, but at the cost of very little generality. In essence, these systems suffer from significant overfitting: to be applicable to a new domain, their rule sets would need extensive or even complete rewriting. Moreover, even for a new dataset within the same domain, for which the schema differs, new rules need to be added in order to accommodate the differences.
    
    In some cases it is beneficial to incorporate a minimal set of rules in KGQA systems, particularly for deriving complex concepts. However, this should be a last resort and not the main translation mechanism, given that a large rule set is hard to maintain and scale.
    
    \item \textbf{Schema-less, incomplete data.} 
    
    One of the strengths of relational databases is to have a database schema which enables strict data modelling and guarantees certain data integrity and data quality aspects. However, since RDF does not strictly enforce a (database) schema, real-world datasets using RDF knowledge graphs often exhibit poor structure \cite{paulheim2013type, kellou2015schema}. Typical examples are properties with missing or generic domains and ranges. In other words, a question answering system over RDF knowledge graphs typically does not have complete schema information. Hence, an important step when working with such incomplete knowledge graphs is to enrich the existing (incomplete) schema, for example, by inferring property ranges and domains based on instance-level data.

\end{itemize}


\section{Bio-SODA: A High-Level Perspective}
\label{Section: Question Answering Pipeline}

In this section we use a motivating example to illustrate the natural language processing pipeline of Bio-SODA. 

\begin{figure}[h!]
    \centering
    \includegraphics[width=\columnwidth]{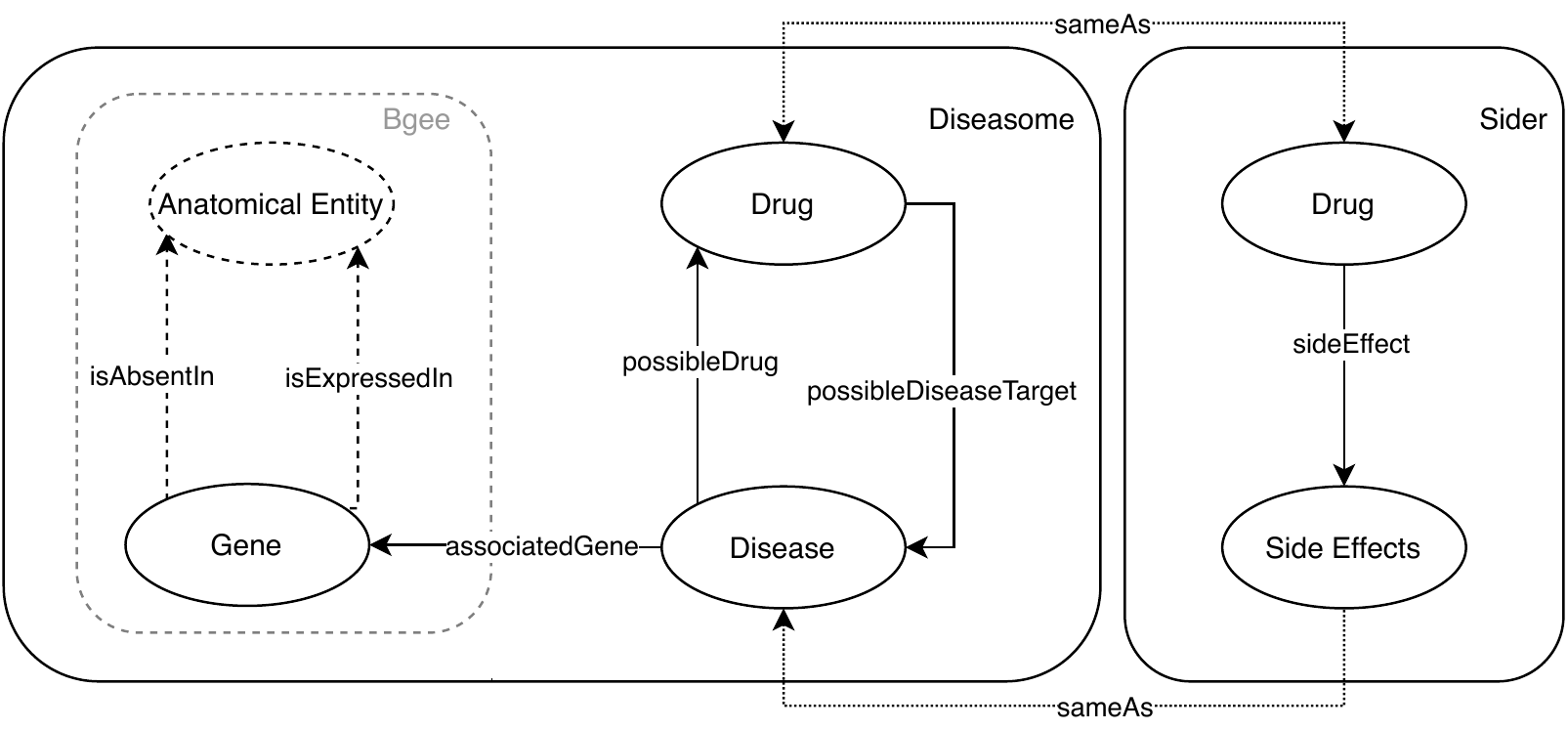}
    \caption{Simplified data model based on the Bgee database and QALD4 \cite{unger2014question} datasets. The data model is a multigraph, including disjoint properties -- such as \textit{isAbsentIn} and \textit{isExpressedIn}, as well as inverse properties, such as \textit{possibleDiseaseTarget} and \textit{possibleDrug}. To make matters more complicated, a \textit{Side Effect} and a \textit{Disease} can be described by the same terms, with instances of the two classes being related via the \textit{sameAs} property. As a result, even simple questions such as \textit{``which drugs might lead to strokes?"} are hard to automatically translate correctly in the absence of external knowledge (\textit{i.e.} ``lead to" = ``side effect").}
    \label{fig:Data Model}
\end{figure}

Consider the data model illustrated in Figure \ref{fig:Data Model}, which combines four different scientific databases. The database \textit{Bgee} on the left contains information about genes and in which parts of the body (anatomical entity) a gene is expressed or absent. The database \textit{Diseasome} in the middle contains information about diseases, as well as drugs targeting each disease. In addition, the drugs are part of the pharmaceutical database \textit{DrugBank} (not explicitly shown in the figure). Finally, the database \textit{Sider} contains information about drugs and their side effects. Correspondences between equivalent drugs in Sider and DrugBank are made through the \textit{sameAs} property. 

Further assume that a domain expert is interested in answering the question: ``\textit{What are the drugs for diseases associated with the BRCA\footnote{Note that, based on the biomedical literature, mutations in the two \textit{BRCA} genes, BRCA1 and BRCA2 (stemming from \textit{BReast CAncer}) are known to be associated with multiple types of cancer.} genes?}". 

\begin{figure}[h!]
    \centering
    \includegraphics[width=\columnwidth]{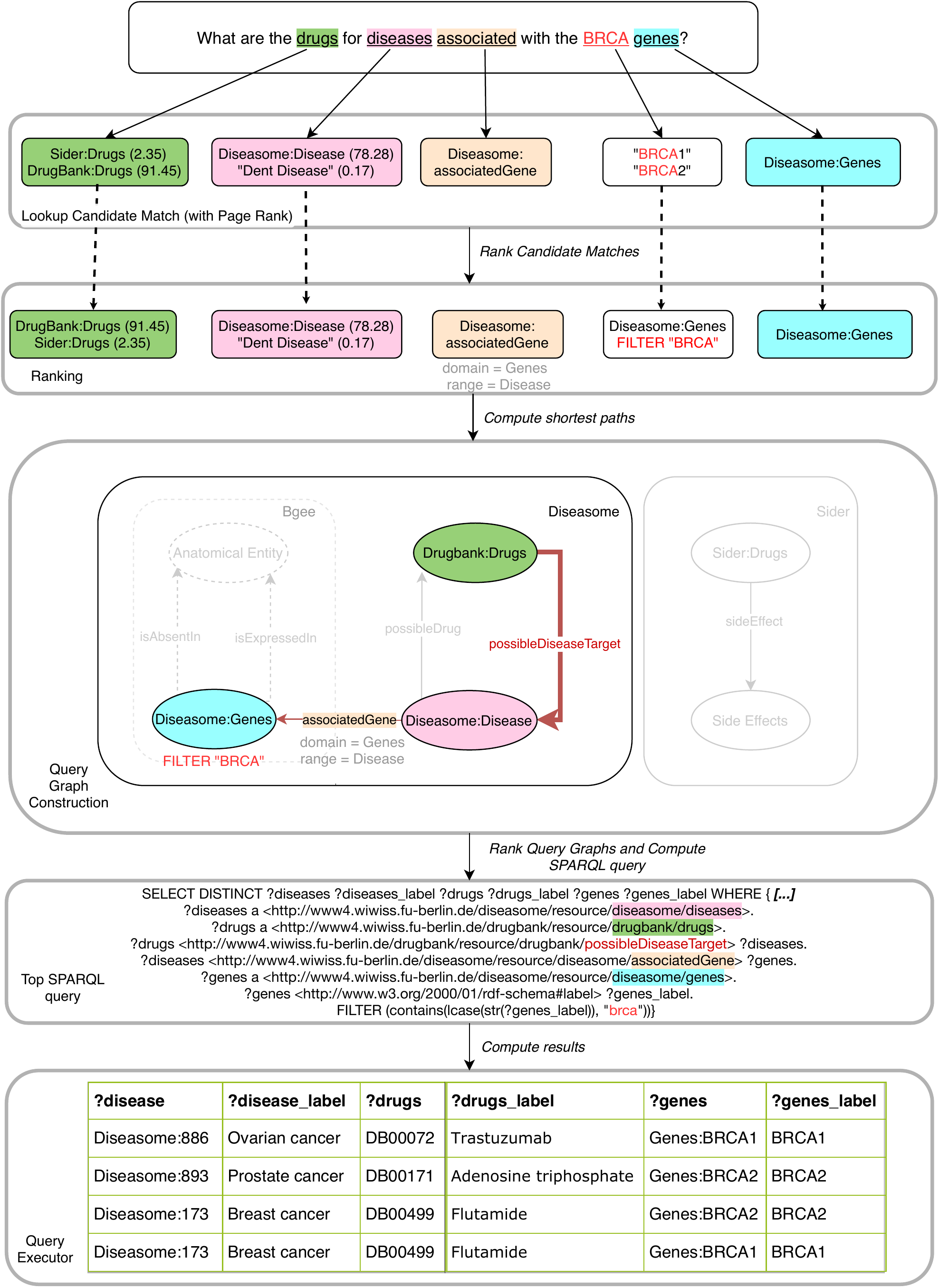}
    \caption{Simplified answer pipeline for the query ``\textit{What are the drugs for diseases associated with the BRCA genes?}". For the sake of simplicity, PageRank scores are solely displayed when more than one match is found.}
    \label{fig:Example Question}
\end{figure}

The natural language processing pipeline of Bio-SODA for answering this question is illustrated in Figure \ref{fig:Example Question}. In particular, the main steps involved in translating the natural language question to SPARQL are as follows: first, Bio-SODA matches question tokens, such as "drugs" and "diseases", against the data stored in the database, using an inverted index. This step is called \textit{Lookup Candidate Match}. In this example, all tokens are of length one, \textit{i.e.} composed of a single word. The inverted index enables retrieving not only the URI of each matching candidate, but also its \textit{PageRank} score. An example is shown in parentheses for the first two tokens in the Figure. In addition, the inverted index retrieves the \textit{class and property names} of the match (omitted in the figure for simplicity). For example, the lookup for ``\textit{BRCA}" retrieves instances of the class \textit{Diseasome:Genes}, where the \textit{rdfs:label} property matches the user token (``\textit{BRCA1}", ``\textit{BRCA2}"). A few simplified Inverted Index entries are provided in Table \ref{tab:inverted_index}. 

In the \textit{Ranking} step, candidates are grouped together according to class/property\footnote{a \texttt{FILTER} for the token \textit{BRCA} is created on the \textit{Diseasome:Genes} class} and ranked according to string similarity and PageRank score. 

In the \textit{Query Graph Construction} step, all the ranked candidates are used to construct a query graph which represents one possible answer or interpretation of the natural language question. For simplicity, Figure \ref{fig:Example Question} only shows the query graph obtained for the top ranked candidate matches. However, Bio-SODA generates multiple alternative interpretations, for example, also including the interpretation considering \textit{Sider:Drugs} instead of the \textit{DrugBank:Drugs}. This can be tested in the demo page of Bio-SODA for QALD4. 

Next, Bio-SODA generates the corresponding SPARQL query for each query graph. Finally, the results are returned by executing the query on the target knowledge graph (see bottom of Figure \ref{fig:Example Question}).


\section{Bio-SODA: System Architecture}
\label{sec:System Architecture}

\begin{figure}[h]
    \centering
    \includegraphics[width=\columnwidth]{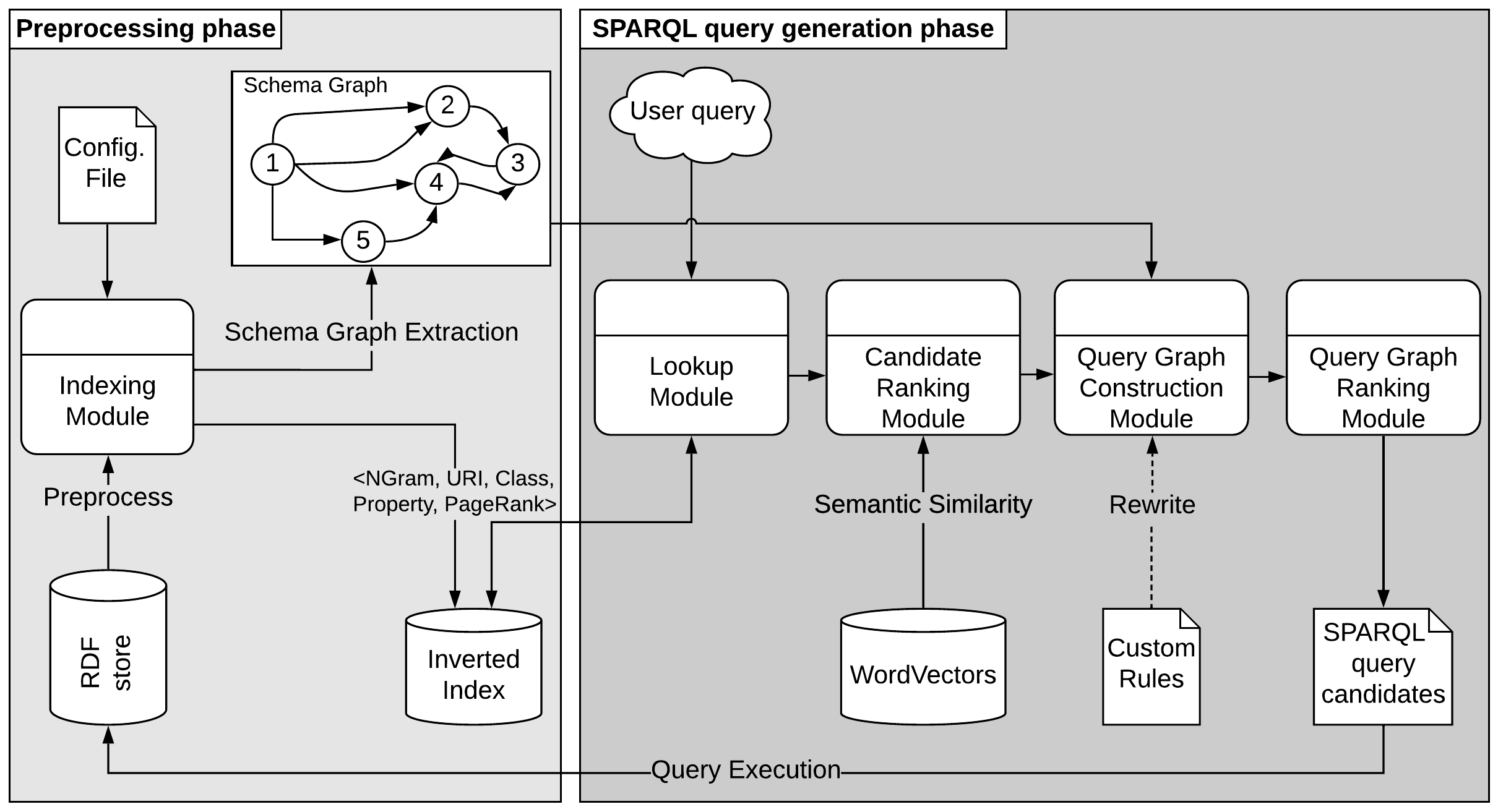}
    \caption{Bio-SODA System Architecture}
    \label{fig:System Architecture}
\end{figure}

The main building blocks of the Bio-SODA system architecture, shown in Figure \ref{fig:System Architecture}, are the following:

\begin{itemize}
    \item {\it Preprocessing Phase}: This phase includes building indexes for efficient lookup as well as automatically generating a schema graph, which will serve as the basis for constructing candidate SPARQL queries in response to user questions. This phase is only executed once, when initialising the system.
    \item {\it SPARQL Query Generation Phase}: This phase represents the natural language query translation process and includes (1) looking up query tokens in the database, (2) ranking the candidate tokens, (3) constructing the candidate query graphs, (4) ranking the query graphs in order of relevance to the user question; and finally (5) constructing a valid SPARQL query and presenting the results. 
\end{itemize}

We will now discuss these phases in more detail.

\subsection{Preprocessing Phase}
The core component of this phase is the {\it Indexing Module}, which extracts the {\it Inverted Index} as well as the {\it Schema Graph} of the RDF data sources:
    \begin{itemize}
        \item {\it Inverted Index}: This index stores the vocabulary of the system. More precisely, all the properties that should be searchable from the RDF data store are indexed, according to a configuration file that specifies the list of properties of interest (by default, all string literals will be indexed). A further configuration option is whether URI fragments should also be parsed and indexed. In this case, these fragments are split by a predefined punctuation list, and through a camel case regex (\textit{e.g.}, ``possibleDiseaseTarget" will be indexed as the corresponding keywords ``possible disease target"). 
        
        The inverted index is stored in a relational database for fast searches and it is used to match tokens (sequences of keywords in a user query) against the RDF data. More precisely, the index stores: keywords (N-grams of literals indexed), the indexed instance URI, the class of this instance, the property from which the keywords were indexed (\textit{e.g.} label), as well as the PageRank score of the instance (see Table \ref{tab:inverted_index}). PageRank scores are computed using the approach presented in \cite{diefenbach2018pagerank}. 
        \\

\begin{table*}[]
\centering
\begin{tabular}{p{0.1\textwidth}p{0.3\textwidth}p{0.2\textwidth}p{0.2\textwidth}p{0.1\textwidth}}
Lookup Key & URI           & Class              & Property             & PageRank  \\ \hline
stroke     & side\_effects:C0038454 & sider:side\_effects & sider:side-EffectName & 0.34     \\
drug       & drugbank:drugs & owl:Class          & rdfs:label           & 91       \\
drug       & sider:drugs    & owl:Class          & rdfs:label           & 2.3   \\
possible disease target &   diseasome:possible-DiseaseTarget & rdf:Property & uri\_match & 80
\end{tabular}
\caption{Inverted Index Sample. The lookup key is used for fast searches based on keywords from a user question. The remaining information is used in attaching candidate matches to the Schema Graph (see description in Section \ref{sec:System Architecture}) in order to construct the corresponding query graphs. A lookup key can consist of multiple keywords. The same lookup key can appear multiple times.}
\label{tab:inverted_index}
\end{table*}

        \item {\it Schema Graph Extractor}: This module is used in order to enrich the (incomplete) schema of the knowledge graph(s) using instance-level data from the RDF store. The Schema Graph is essentially the {\it accurate schema of the integrated RDF data} which Bio-SODA automatically extracts from data instances\footnote{Note that multiple RDF sources can be combined, as long as they are semantically aligned - \textit{i.e.} they have at least one common concept, such as \textit{Gene}.}. Moreover, the Schema Graph serves as the basis for constructing candidate query graphs from selected entry points (\textit{i.e.},  matches  for  tokens in  a user question). 
        
        Computing a Schema Graph allows the system to compensate for incomplete schema information, for example, in cases where domains and ranges for properties are either missing or ill-defined. A second benefit of the Schema Graph is that it enables integrating multiple data models from different knowledge graphs. 
        
        Extracting the schema graph is achieved via SPARQL queries that compute, for example, domains and ranges of all properties, based on the classes of the instances which they connect. As a simplified example, a triple asserting ``Migraine $\rightarrow$ possibleDrug $\rightarrow$ Ibuprofen" will result in \textit{Disease} $\rightarrow$  \textit{possibleDrug} $\rightarrow$ \textit{Drug} being added to the Schema Graph. 
        
        Currently, as a minimum requirement we assume that each instance in the RDF data has a well-defined class, \textit{i.e.} an explicit \textit{rdf:type}. If this is not the case, additional preprocessing with external tools (for example, using RDF schema discovery techniques \cite{kellou2015schema}), would be required in order to properly define types for all RDF instances.
    \end{itemize}
  
  We note here that indexing is a preprocessing step that is only required once, when the system is initialized. Afterwards, updates to the RDF store can be incorporated periodically through incremental updates (appends) to the inverted index, while the Schema Graph only needs to be recomputed in case of schema changes.

\subsection{SPARQL Query Generation Phase}

Given a natural language question, the goal of the Bio-SODA system is to translate it into a set of ranked candidate SPARQL queries, such that the top ranked query is the closest to the user's query intent. In the following, we detail the role of each component involved in this translation process, namely the {\it Lookup Module}, the {\it Candidate Ranking Module}, the {\it  Query Graph Construction Module}, the {\it Query Graph Ranking Module} and the {\it Query Executor Module}.
 
  \begin{itemize} 
    \item {\it Lookup Module}:
    
    The lookup module has the role of retrieving the best candidate matches for tokens identified in a user query. A token is defined by the longest sequence of keywords that matches an entry in the Inverted Index (implemented in a relational database for fast searches). For example, in the question ``\textit{What are the possible disease targets of Ibuprofen?}" the two tokens extracted will be ``\textit{possible disease target}" (corresponding to an RDF property name) and ``\textit{Ibuprofen}" (corresponding to one or more Drug instances).
    \\
    
    \item {\it Candidate Ranking Module}:
    
    The lookup module can return a large number of candidate matches per token. In order to find best candidate matches, the ranking module {\it groups together equivalent matches and ranks them} in order of relevance to the initial query. 
    For example, instances of the class \textit{Drug} with matching \textit{rdfs:label}  are grouped together. In our running example illustrated in Figure \ref{fig:Example Question}, the genes \textit{BRCA1} and \textit{BRCA2} are a match for the keyword \textit{BRCA}. 
    
    Furthermore, both {\it string similarity} and {\it node importance} are taken into account when ranking. Including the PageRank score as a measure of importance in the knowledge graph reduces the influence of the quality of labels assigned (labels which can be imprecise, see discussion in Section \ref{Problem Statement}). 
    
    The intuition behind this is that domain knowledge graphs usually cluster around a few important concepts, which will be reflected in the PageRank scores of the corresponding nodes. For example, UniProt\footnote{\url{https://sparql.uniprot.org/}} \cite{redaschi2009uniprot}, a protein knowledge base containing more than 60 billion triples, includes only 177 classes at the time of writing. Out of these, only few classes, such as \textit{Protein} and \textit{Annotation}, have a central role, and will usually be the target of domain expert questions. 
    
    Likewise, in the case of the CORDIS EU projects dataset (see Section \ref{Evaluation} for details), two different classes of Projects are available, \textit{EC-Project} and \textit{ERC-Project}. However, there is significantly more information in the dataset for the first class. In the lack of query logs or handcrafted rules for mapping query tokens to the correct candidates, the PageRank score can serve as a good proxy for ranking candidates according to node centrality, similarly to the initial approach used by web search engines \cite{page1999pagerank}. 
 
    As an added benefit, scoring with PageRank also ensures that metadata matches are prioritized. For example, \textit{Drug} as a class name will rank higher than an instance match.
    
    Finally, to ensure that candidate matches not only have good string similarity, but are also {\it semantically similar}, word embeddings are also used in the candidate ranking. The similarity comparison ensures that spurious matches, such as \textit{gene} compared to \textit{oogenesis}, are discarded based on a pre-defined similarity threshold in the system configuration. 
    
    Any word embeddings can in principle be used with Bio-SODA. For the two main bioinformatics use cases considered in this paper, we use Word Vectors extracted from PubMed, as described in \cite{moen2013distributional}. 
    The candidate ranking module presents to the user top N matches per query token, where N is configurable in the system.
    \\

    \begin{algorithm}[ht!]
    \KwData{\\$M^{n\times t}$: the matrix of ranked candidate matches, where\\
    \ \ $n$ = the number of candidate matches per token, \\
    \ \ $t$ = the number of tokens in the user question. \\ 
    $M_i$  = a set of candidates covering one match per token (i.e.   the $i^{th}$ row vector of the $M^{n\times t}$ matrix).\\
    $G$: Schema Graph of the RDF data}
    \KwResult{$S$: the ranked set of candidate query graphs}
    
    \ForEach{ $M_i \in M$ }{
        $QG_i = \phi$ (empty graph) \\
        \ForEach{ candidate match $T_j \in M_i$ }{
            \eIf{$T_j$ = a RDF property}{
            Get domain $D$ and range $R$ of $T_j$ from $G$; \\
            Add $D$ and $R$ as vertices to $QG_i$; \\
            Add edge $T_j$ between $D$ and $R$ in $QG_i$; \\
            \textbf{if} multiple domains / ranges for $T_j$ \textbf{then} Create a new copy of $QG_i$ per alternative; }
            {
            Compute in schema graph $G$:\\
            \ \  shortest paths between class of $T_j$ and classes of other matches $T_z$ in $M_i$;\\
            Add shortest paths to $QG_i$ \\
            \textbf{if} multiple alternatives exist \textbf{then} \\ Create a new copy of $QG_i$ per alternative; 
            }
        } 
        Add spanning tree extracted from $QG_i$ to result set $S$ (Steiner tree approximation)
    } 
    $S\_sorted$ = sort $S$ by sum of match score of composing vertices. On a tie, sort by the weight (i.e. the number of edges) of spanning tree.\\
    \textbf{return} $S\_sorted$
    \caption{Iterative graph-based approach for constructing query graphs from candidate matches}
    \label{algo:Query Graphs Construction}
\end{algorithm}

    \item {\it Query Graph Construction Module}:
    
    The goal of this module is to use the matches from the previous step to {\it generate a list of candidate query graphs}. We extend the approach presented in \cite{gkirtzou2015keywords} to translate matches to query graph patterns. More precisely, we apply the iterative algorithm shown in Algorithm \ref{algo:Query Graphs Construction}: for each set of candidate matches (one match per query token), we augment the Schema Graph by attaching the candidate matches to their corresponding class. Next, we find the minimal subgraph that covers all matches. For this purpose, we solve the approximate Steiner tree problem by computing the minimal spanning tree that covers one match per token. 
    
    Note that there might be multiple such subgraphs, given that two classes can be connected via multiple properties. However, unless the user can be involved in disambiguating, it is important to generate all the variants, given that two equal-length subgraphs might actually have opposite semantics. Recall the example shown in Figure \ref{fig:Data Model}, where the properties \textit{e.g}, \textit{isAbsentIn} \textit{versus} \textit{isExpressedIn} both connect the same two classes, but represent disjoint result sets.


    Finally, in some cases handcrafted rules for inferring new concepts or relationships are required, due to the complexity of the corresponding query graphs. In such cases translating user questions into SPARQL cannot be done via simple entity linking methods. Therefore, if needed, our approach also supports adding rules to derive implicit information from the original knowledge graph as part of the question answering pipeline. These rules are implemented as sub-queries similar to the \texttt{SELECT} SPARQL query form. In this case, the rule head is the SPARQL query projection, and the rule body is the WHERE clause content.  
    \\
    


    \item {\it Query Graph Ranking Module}:

    The query graph ranking module plays an important role in presenting the user with a meaningful, ordered list of results. In contrast to existing work, we do not return the {\it overall minimal subgraph} as the top result, but rather the {\it graph that maximizes the sum of the match scores} of the candidates covered. To understand why this is the case, consider the following  question: \textit{``What are the drugs for asthma?"}. This question translates to a 2-hop query graph, joining \textit{Drug} and \textit{Disease} via the \textit{possibleDiseaseTarget} path (see Figure \ref{fig:Data Model}). However, one likely scenario is that the description of a \textit{Drug} instance includes the keyword \textit{asthma}. In this case, the minimal query graph would be 1-hop only, retrieving only \textit{Drug} instances that explicitly contain the keyword in the description, probably a small subset of all instances which have the corresponding \textit{Disease} as a possible target. In this case, the minimal result would have good precision, but very low recall.
    \\

    \item {\it Query Executor Module}:
    
    Finally, the query executor translates the ranked query graphs into SPARQL queries, assigning meaningful variable names, also adding human-readable fields to the result set wherever possible. Importantly, we do not only return the best result, but rather a ranked list of possible interpretations (top N, where N is configurable in the system). This gives the user the opportunity to inspect the results in order to choose only the interpretation (\textit{i.e.} SPARQL query) that matches the question intent. 
\end{itemize}


\section{Experiments}
\label{Evaluation}

In this section we evaluate the F1-score performance of Bio-SODA for translating natural language questions to SPARQL and compare it against state-of-the-art systems for querying RDF-based knowledge graphs. Note that we focus on top-performing open-source systems that are publicly available for testing and do not require training data \cite{affolter2019comparative}. 

In particular, we tested Sparklis \cite{ferre2017sparklis}, a generic query builder system for knowledge graphs\footnote{A live demo can be tested with any SPARQL endpoint at \url{http://www.irisa.fr/LIS/ferre/sparklis/}}.  Furthermore, we compared against GFMed \cite{marginean2017question} which was top ranked in the QALD4 biomedical challenge and specifically designed for this dataset. Apart from this, we use GFMed's publicly available grammar\footnote{See \url{http://cs-gw.utcluj.ro/~anca/GFMed/index.html}} to evaluate how the system performs outside of the official QALD4 biomedical dataset. In addition, we compared our approach against SQG \cite{zafar2018formal}, a system for query generation over knowledge graphs\footnote{Available at \url{https://github.com/AskNowQA/SQG/}}.

\subsection{Datasets}

\begin{table*}[h!]

\footnotesize
\begin{tabularx}{\textwidth}{|X|c|r|r|r|}
\hline
Dataset          & Sources                    & \#Classes & \#Triples & Size on Disk  \\ \hline
QALD4-biomedical & Drugbank, Diseasome, Sider & 12       & 0.69 M    & 200 MB \\
Bioinformatics   & Bgee, OMA                  & 37       & 430 M    & 30 GB \\
CORDIS           & EU projects dataset        & 26       & 6.5 M    & 1 GB   \\

\hline
\end{tabularx}
\caption{Descriptions of the 3 public datasets used in our evaluation.}
\label{tab:dataset}
\end{table*}

Three datasets were considered for evaluating Bio-SODA, see Table \ref{tab:dataset}. Importantly, all three are real-world, in-use datasets. For each dataset, we briefly highlight the specific challenges that need to be tackled in the context of designing a generic question answering system: 
\begin{enumerate}
    \item The \textit{QALD4 biomedical} dataset is composed of Sider, DrugBank and Diseasome. This dataset includes several challenges such as multiple \textit{Drug} classes and identical terms describing both \textit{Disease}  and \textit{Side Effects} instances, which are connected via \textit{owl:sameAs} properties. 
    \item The \textit{bioinformatics} dataset is composed of the Bgee (gene expression) \cite{bastian2021bgee} and OMA (orthology) \cite{altenhoff2021oma} RDF stores. Given the highly specialized domain information contained in these sources, a particularity of this dataset is that questions can include complex concepts which translate to long SPARQL query graphs. An added challenge deriving from this is that the same concepts can be connected through multiple equal-length paths with semantically different or even opposite meanings. 
    \item The \textit{CORDIS} dataset of EU-funded projects. Although this dataset has a simpler schema, the challenge here is that questions can have a higher degree of ambiguity. In some cases, multiple interpretations are valid -- for example, many terms are reused often and in a variety of contexts, such as ``\textit{Big Data}". This can be either part of a project title, a topic or even an organization name. Therefore, identifying the query intent in some cases (\textit{e.g.} \textit{Show Big Data projects}) cannot be done without user disambiguation.
\end{enumerate}

\subsection{Queries}


We have reused the official 50 queries of the QALD4 biomedical challenge\footnote{\url{https://github.com/ag-sc/QALD/blob/master/4/data}}. We do not distinguish between training and test queries. Indeed, we report performance metrics for all systems we tested across the entire set of 50 queries. Given that the test set was also available to participants in the official challenge, we believe this to be a fair evaluation. We do not change the questions in the official challenge, not even in cases where we could identify mistakes in the question. Furthermore, as opposed to previous work using this benchmark \cite{song2015tr}, we do not materialize triples based on \textit{owl:sameAs} statements and only use the exact dataset, as provided in the official benchmark.

For the bioinformatics dataset, in collaboration with domain experts, we created a benchmark of 30 queries, in increasing order of complexity, across two datasets, namely Bgee and OMA. The queries represent real information needs of domains experts within the field of gene expression and orthology, using the publicly available RDF data of Bgee\footnote{\url{https://bgee.org/sparql}} and OMA\footnote{\url{https://sparql.omabrowser.org/sparql}}. The average number of triple patterns per query here is 7 (not taking into account joint queries between the two sources, which have even higher complexity), with some questions jointly targeting 4 entities or more (\textit{Gene}, \textit{Species}, \textit{Anatomical Entity}, \textit{Developmental Stage}). In contrast, in existing benchmarks, such as LC-Quad \cite{trivedi2017lc}, queries with only 2 entities are already considered complex.

In order to test Bio-SODA using an entirely different domain, using the CORDIS dataset of EU funded projects, we created a test set of 30 queries in increasing order of complexity. Given the relatively simple structure of this data model, the average number of triple patterns per query is close to that of existing KGQA benchmarks \cite{trivedi2017lc}, with an average 2.3 triple patterns per query. However, the complexity stems from the usage of filters, literals in the query, as well as the higher degree of ambiguity.

Queries across the three datasets include aggregations, negations, and make extensive use of filters.

All questions, as well corresponding SPARQL queries, are available in the Evaluation folder of our GitHub repository\footnote{Evaluation in \url{https://github.com/anazhaw/Bio-SODA/}}.



\subsection{Results}

We use the standard evaluation metrics of precision (P), recall (R) and F1-score, macro-averaged over all questions in the dataset. For Bio-SODA in particular, although the system generates a ranked list of possible interpretations, we report numbers based on the top answer only (Precision@1). The results are presented in Table \ref{tab:results} and discussed in the following section. 
For easy accessibility to the Bio-SODA system, as well as reproducibility of the results, we also provide a demo page for each of the three datasets, available online (see Section \ref{sec:Introduction}).


\begin{table}[h]
\captionsetup{width=\columnwidth}
\begin{tabularx}{\columnwidth}{|X|r|r|r|}
\hline
\textbf{Datasets and Systems} & \textbf{Precision} & \textbf{Recall} & \textbf{F1}  \\ \hline
\hline
\textbf{Dataset 1: QALD4}        &  &  &    \\ \hline
GFMed                 & 1      & 0.99    & 0.99 \\
SQG                   &   0.42   & 0.42  &  0.42  \\
Sparklis (5.5 steps/query)             &   0.88   & 0.88     &  0.88   \\
Bio-SODA              &    0.61       &   0.60     &  0.60  \\ \hline
\hline
\textbf{Dataset 2: Bioinformatics} &           &        &    \\ \hline
GFMed                 & 0      & 0    & 0 \\
SQG                   &   0.16        &  0.16      & 0.16  \\
Sparklis              &    -         &   -       &  -   \\ 
Bio-SODA              &     0.6      &  0.6      &  0.6  \\  \hline
\hline
\textbf{Dataset 3: CORDIS}         &           &        &     \\ \hline
GFMed                 & 0      & 0    & 0  \\
SQG                   &  0.33         &  0.33      &  0.33  \\
Sparklis  (6.2 steps/query)            &  1         &   1     & 1  \\ 
Bio-SODA              &   0.66        &   0.66     &  0.66 \\ \hline

\end{tabularx}
\caption{Performance of translating natural language questions to SPARQL. By considering a perfect user of the Sparklis tool, the minimum number of manual steps for composing a query (averaged over all queries) is shown between parentheses. }
\label{tab:results}
\end{table}

We will now discuss the performance of each system in more detail.

\textit{GFMed} shows the highest F1-score for the QALD4 dataset. However, it cannot (nor was it intended to) be used outside this dataset without rewriting the set of grammar rules that are strictly designed for question answering over specific releases of \textit{Diseasome}, \textit{Drugbank} and \textit{Sider}. Hence, the F1-score for the bioinformatics dataset and the CORDIS datasets is 0.

\textit{SQG} on the other hand, originally evaluated on the LC-Quad \cite{trivedi2017lc} benchmark, does not support complex multi-hop questions, nor filters or queries involving literals. ``\textit{Show me projects which started in \textbf{2020}?}" is an example of such a query, where \textit{2020} is a numerical literal, as opposed to a linkable entity. While in the case of LC-Quad these limitations do not impact performance, all three datasets considered in our evaluation include such features, which explains the poorer performance of SQG: an F1-score of 0.42 in the case of QALD4, only 0.33 in the CORDIS dataset, and finally 0.16 in the case of the bioinformatics dataset. We note that these results are a theoretical best, since for SQG we assume perfect entity and property linking, leading to the highest performance it can achieve.  

Finally, \textit{Sparklis} is not a question answering system per-se, but rather a query builder, which helps users form the correct question by composing building blocks starting from examples of class names, properties, values etc. Therefore, in order to answer questions, we needed to rephrase them from the available building blocks  \textit{manually}. On the positive side, we found Sparklis to be a powerful system, because it enables building a rich variety of query types out-of-the-box. To achieve this, only the SPARQL endpoint URL of the target RDF data store is required. 

Using the query building methodology of Sparklis, 44 out of 50 questions in the QALD4 biomedical benchmark can be answered. Furthermore, all questions in the CORDIS dataset can also be answered. Although this result might seem surprising, recall that the major challenge of this dataset is disambiguation. The manual query building process in Sparklis addresses exactly this problem, provided that the user knows very well how the data are structured and semantically represented. Therefore, on the negative side, we found that the query building methodology requires precise understanding of the data model, especially if multiple classes have the same label, as is the case in QALD4. 

For example, answering the question \textit{Which drugs might lead to strokes?} requires knowing that the \textit{Drugs} class to be used is the one in \textit{Sider}, as opposed to the one in \textit{Diseasome}. Furthermore, formulating questions in Sparklis is a manual and therefore time-consuming process. Even when making the strong assumption that the user has perfect knowledge of the data model, as well as of the features of Sparklis (for example, how to correctly formulate aggregations, which can be challenging), the minimal number of manual steps required to formulate questions is on average 5.5 interactions per question for QALD4 and 6.2 for CORDIS, with a maximum of 10 for the more complex questions. In most cases, the question resulting from composing the building blocks will be significantly different from a true natural language question. 

We did not pursue this approach on the bioinformatics dataset, because complex concepts in this dataset (ortholog, paralog) cannot be expressed through the query building mechanism. More precisely, Sparklis does not support complex property paths. 

\textit{Bio-SODA} is a middle-ground between the generic, but manual approach of Sparklis, and the grammar-based approach of GFMed, which is not easily transferable to a new domain. More precisely, Bio-SODA achieves relatively good performance (around 0.6 F1-score) across the three datasets without requiring manual intervention. The only exception are two custom rules for the bioinformatics dataset, which help answer 4 out of 30 queries. 

Although GFMed has the best results for QALD4, it cannot be used outside this dataset without a complete rewriting of the grammar rules. Sparklis also achieves better results on the two datasets tested, but with the big disadvantage that it is a manual approach, where the user must understand the data model in order to compose questions correctly. Our findings are further detailed in the Evaluation folder in our GitHub repository.  

\subsection{Impact of Ranking Algorithm}

In this section we study the impact for our ranking algorithm on the performance of Bio-SODA. In particular, we conducted an ablation study to quantify the importance of ranking by PageRank score of candidate matches. For this purpose, we disable our ranking algorithm and instead use a simple string similarity-based ranking algorithm for candidate matches, returning the \textit{overall} minimal subgraph as the top answer. 

The results, displayed in Table \ref{tab:ablation}, show that ranking makes a crucial difference, in particular for the CORDIS dataset. The reason for this is that for most of the keywords that describe metadata (such as class names, like \textit{Project Topic} or \textit{Subject Area}), there exists in the dataset a project whose acronym matches exactly. For example, there exist projects with acronyms such as \textit{Topic}, \textit{Area}, \textit{Host}, \textit{Code}, which are (according to string similarity only) classified as best matches for  tokens in the original question. Constructing the \textit{overall} minimal subgraph leads to wrong results in almost all cases, except for only 3 out of 30 questions, where there is no ambiguity. Note that adding \textit{no other change} aside from considering PageRank scores in ranking enables answering 17 more queries out of 30 for this dataset.

\begin{table}[h]
\begin{tabularx}{\columnwidth}{|l|>{\raggedleft\arraybackslash}X|>{\raggedleft\arraybackslash}X|}
\hline
\textbf{Dataset} & \textbf{(a) Correct with Bio-SODA Ranking} & \textbf{(b) Correct with String Similarity Ranking} \\ \hline
QALD4            & 30/50       &   23/50       \\ \hline
Bioinformatics   & 18/30       &   12/30          \\ \hline
\textbf{CORDIS}           & \textbf{20/30}       &    \textbf{3/30}                              \\ \hline
\end{tabularx}
\caption{Ablation study on the Bio-SODA performance of translating natural language questions to SPARQL: (a) SPARQL candidate query ranking with node centrality measure versus (b) traditional ranking approach with string similarity and overall minimal subgraph as top result.}
\label{tab:ablation}
\end{table}

\subsection{Error Analysis and Remaining Problems}
\label{sec:error_analysis}

In the QALD4 biomedical benchmark, Bio-SODA correctly answered 30 out of 50 questions with an additional 2 partially correct. We note that 1 question in QALD4 cannot be answered by Sparklis nor Bio-SODA due to missing label information. More precisely, the instance \url{<http://www4.wiwiss.fu-berlin.de/diseasome/resource/genes/EDNRB>} is the target of the question ``Which genes are associated with Endothelin receptor type B?". However, the label \textit{Endothelin receptor type B} is not assigned in the official dataset of the benchmark, nor can it be derived from the URI fragment, for example. Upon closer inspection, it becomes clear that the question is ill-formulated. Since \textit{EDNRB} itself is a gene, the correct question should be ``Which \textit{diseases} are associated with EDNRB?". In total, we have found at least 4 out of 50 entries in the dataset to contain errors, either in the question formulation, or in the ground truth answer. These have already been discussed in previous studies \cite{song2015tr}.

\begin{figure*}[h]
    \centering
    \includegraphics[width=1.3\columnwidth]{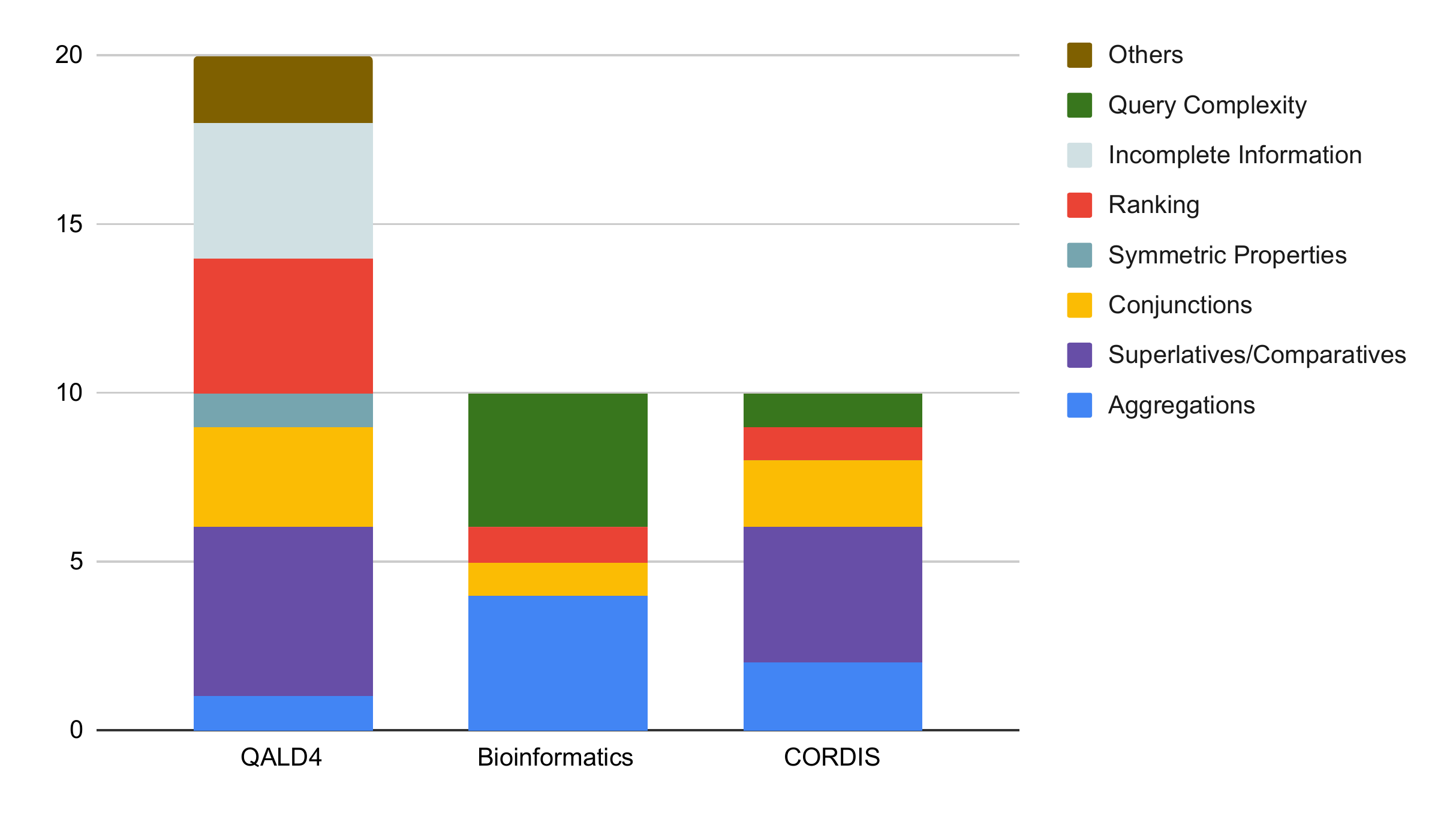}
    \caption{Bio-SODA failure analysis. Out of the total 50 questions in the QALD4 biomedical benchmark, Bio-SODA cannot correctly answer 20. A further 12 out of 30 cannot be answered in the bioinformatics dataset, mainly due to query complexity (some queries having more than 10 triple patterns). Finally, on the CORDIS dataset 10 out of 30 queries cannot be answered, a large fraction of which include features currently unsupported in Bio-SODA: aggregations, comparatives, conjunctions etc.
    }
    \label{fig:Failure_Analysis}
\end{figure*}

An additional number of questions cannot be answered by Bio-SODA across the three datasets due to other reasons. We summarise them in Figure \ref{fig:Failure_Analysis}, explained in the following:
\begin{itemize}
    \item{\it Aggregations}. Our system currently does not support questions that require aggregations, such as \textit{Count}, \textit{Sum} etc. An example of such a question would be \textit{Count the projects in the life sciences domain}. A possible solution to this would be to include pre-defined patterns or training a question classifier for this purpose.
    \item {\it Superlatives/Comparatives}. Another unsupported feature in the current prototype is the use of quantifiers (superlatives or comparatives). An example would be \textit{Which drug has the highest number of side-effects?} 
    \item {\it Conjunctions}. Conjunctive questions which involve multiple instances of the same class are not supported in the current prototype. An example of such a case is \textit{List drugs that lead to strokes and arthrosis}. This limitation derives from our methodology in computing the minimal subgraph covering candidate matches, which would require special handling for cases when multiple candidates of the same class are present in a question.
    \item {\it Properties with same domain and range}. Stemming from the same limitation mentioned above, these properties are currently not supported. In QALD4, the only instance of this is the \textit{diseaseSubtypeOf} property, which has the \textit{Disease} class as both domain and range. In the bioinformatics dataset we handle symmetric properties describing \textit{ortholog} and \textit{paralog} genes through custom rewrite rules.
    \item {\it Ranking}. One of the major sources of failure in our prototype remains ranking. In the QALD4 dataset, ranking problems affect 4 out of 50 queries. An example is: \textit{What are the diseases caused by Valdecoxib?}. Here, the system cannot correctly choose \textit{Drug - sideEffect - Side\_Effect} over the alternative \textit{Disease - possibleDrug - Drug}. The reason for this is that the \textit{Disease} class matches exactly the term in the question, while the \textit{Drug} class in \textit{Diseasome} has a higher PageRank score than the one in \textit{Sider}. 
    
    \item {\it Incomplete information}. This problem affects mainly the results in the QALD4 dataset, more precisely 4 out of 50 queries. We have already covered the example of the question targeting the \textit{EDNRB} gene, which lacks the correct label in the official dataset. We currently do not enrich the inverted index with synonyms or external information, which means questions must be formulated in terms of the available vocabulary of the dataset. However, this limitation could be addressed by indexing synonyms from external data sources. Additional three questions cannot be answered because they refer to URIs that do not have any class defined in the data, therefore the system cannot attach the candidate matches anywhere in the Schema Graph. 
    
    An example is the \textit{drugType} property, which can take two values, either \textit{\url{http://www4.wiwiss.fu-berlin.de/drugbank/resource/drugtype/experimental}} or \textit{\url{http://www4.wiwiss.fu-berlin.de/drugbank/resource/drugtype/approved}}. We believe a better modelling of the data should have provided, for example, either these as a \textit{xsd:anyURI} datatype, given they are not used for any other purposes, or defined some class for both. 
    \item {\it Query complexity (difficult queries)}. The bioinformatics dataset covers queries with high complexity, which are difficult to solve especially since they include symmetric properties, with multiple instances of the same class, each filtered according to different conditions. 
    
    An example of such a question is: \textit{Retrieve Oryctolagus cuniculus' proteins encoded by genes that are orthologous to Mus musculus' HBB-Y gene}. Here, the task is to retrieve \textit{Gene} instances in a particular \textit{Taxon} (species), namely the rabbit (\textit{Oryctolagus cuniculus}), which are \textit{orthologs} (symmetric property) of a second instance of \textit{Gene}, labeled \textit{HBB-Y}, in a different species, namely the mouse (\textit{Mus musculus}). The resulting query has over 15 triple patterns, with 3 filters (the 2 species names plus the gene name). 
    \item {\it Others}. Two questions in the QALD4 dataset have particular challenges, the first being a stemming error. In the question \textit{Give me drugs in the gaseous state}, the term \textit{gaseous} cannot be correctly stemmed to \textit{gas}. The second type of error is due to unsupported \textit{ASK} queries, \textit{e.g.} \textit{Are there drugs that target the Protein kinase C beta type?}. Here, Bio-SODA retrieves examples of such drugs, instead of the boolean \textit{True}. However, we do not consider this a fundamental limitation and a question type classifier could be added in future work.
\end{itemize}
We report a more detailed analysis of all systems considered in this paper in the \href{https://github.com/anazhaw/Bio-SODA/tree/master/Evaluation}{Evaluation folder} in our GitHub repository. 
%
%

\section{Lessons Learned}
\label{sec:lessons_learned}


Considering the challenges of question answering over knowledge graphs introduced in Section \ref{Problem Statement}, we highlight the following design goals for natural language processing engines:

\begin{itemize}
    \item {\it Generality}: The system should be easily adaptable to new datasets. In particular, the system should be able to answer questions in a new domain with minimal manual intervention and without relying on extensive training data, which is hard to obtain in many domains.
    Along this line, a desirable property is also the ability to cope with ``real-world" datasets, dealing with incompleteness in the data, for example in the form of:
        \begin{itemize}
            \item missing schema information (should be inferred from instance-level data);  
            \item missing labels (should be incorporated from URIs whenever meaningful);
        \end{itemize}
    \item {\it Extensibility}: The system should easily work with multiple datasets (provided they are already semantically aligned---\textit{i.e.}, data integration is a prior requirement). Many studies introduce possible approaches for data integration, including a recent approach for ontology-based data integration, covering one of the bioinformatics use cases presented in this paper \cite{sima2019enabling}.
    
    \item {\it Configurability}: The database owner must be able to specify which properties (\textit{e.g.} labels, descriptions) should be searchable using the system. Our experience with real-world datasets showed that in general it is not desirable for \textit{all} properties to be indexed and thus be searchable. As an example, in many cases, fields in the queried data sources can be either redundant or too verbose. In bioinformatics, these are abstracts of papers that are assigned as values to an RDF property, whose length can therefore be up to 300 words. Similarly, in the CORDIS dataset, these are the abstracts of the EU projects. These cases should be handled through a dedicated approach, for example, based on classical information retrieval methods as discussed in \cite{nadig2020}.
    
    \item {\it Explainability}: The system should clearly guide the user through how a question was processed and interpreted. This starts from explaining which concepts were matched in relation to the original question, continuing with how these candidate matches are composed together in a query graph in order to provide the final SPARQL query. Finally, the query results should be understandable as well. Therefore, the projected variable names should also be meaningful. 
\end{itemize}

\section{Conclusions and Outlook}
\label{Conclusions}

In this paper we have introduced Bio-SODA, a question answering system for domain knowledge graphs, which we evaluated across three real-world datasets pertaining to different domains: biomedical, gene orthology and gene expression, and finally EU-funded projects. Our results have shown that Bio-SODA outperforms state-of-the-art systems that are publicly available for testing by a 20\% F1-score improvement and more. The main advantage of Bio-SODA over existing open-source systems is that it can handle {\it complex, multi-triple pattern queries} without requiring user guidance and training data. Bio-SODA uses a novel ranking approach that takes into account both string and semantic similarity, as well as node centrality of candidate matches. Our experiments demonstrate that our ranking approach improves the quality of results, particularly in the context of datasets which can suffer from redundancy and imprecise labels.


As a first step in future work, we plan to add user feedback to the question answering process by involving the user in a disambiguation dialog for selecting the best candidate matches. We also plan to consider the users' feedback for ranking the best answer among resulting candidate queries. As a long term direction for future research, we envision compiling a benchmark of cross-domain question-answer pairs, similarly to the Spider benchmark in the relational database world \cite{yu2018spider}, which would enable research into refining pre-trained KGQA models for new domains.

\begin{acks}
We thank the Swiss National Science Foundation for funding (NRP 75, grant 407540\_167149), Lukas Blunschi for the implementation of the SODA system for keyword search system over relational databases \cite{blunschi2012soda}, on which our prototype is based, and Katrin Affolter for important contributions to the natural language processing pipeline in Bio-SODA.
\end{acks}
\bibliographystyle{ACM-Reference-Format}
\bibliography{document}

\end{document}